\documentclass[11pt,fleqn]{article}
\usepackage{epstopdf} 
\usepackage{graphicx}
\usepackage[a4paper,left=2cm,right=2cm]{geometry}
\usepackage{amsmath}
\usepackage{float}
\usepackage[latin1]{inputenc}
\usepackage{amsmath}
\usepackage{amsfonts}
\usepackage{amssymb}
\usepackage{float}
\usepackage[toc,page]{appendix}
\floatstyle{boxed} 
\restylefloat{figure}

\newcommand{\be}{\begin{equation}}
\newcommand{\ee}{\end{equation}}
\newcommand{\bea}{\begin{eqnarray}}
\newcommand{\eea}{\end{eqnarray}}

\begin{document}
\title{Phenomenology of $U(1)_{F}$ extension of inert-doublet model with exotic scalars and leptons.}
\author{Lobsang Dhargyal\footnote{Email : dhargyal@hri.res.in} \\\\\ Harish-Chandra Research Institute, HBNI, Chhatnag Road, Jhusi Allahabad 211 019 India.}

\maketitle
\begin{abstract}

In this work we will extend the inert-doublet model (IDM) by adding a new $U(1)_{F}$ gauge symmetry to it, under which, a $Z_{2}$ even scalar ($\phi_{2}$) and $Z_{2}$ odd right handed component of two exotic charged leptons $(F_{eR},\ F_{\mu R})$, are charged.  We also add one $Z_{2}$ even real scalar ($\phi_{1}$) and one complex scalar ($\phi$), three neutral Majorana right handed fermions ($N_{1},\ N_{2},\ N_{3}$), two left handed components of the exotic charged leptons $(F_{eL},\ F_{\mu L})$ as well as $F_{\tau}$ are all odd under the $Z_{2}$, all of which are not charged under the $U(1)_{F}$. With these new particles added to the IDM, we have a model which can give two scalar DM candidates, together they can explain the present DM relic density as well as the muon (g-2) anomaly simultaneously. Also in this model the neutrino masses are generated at one loop level. One of the most peculiar feature of this model is that non-trivial solution to the axial gauge anomaly free conditions lead to the prediction of a stable very heavy partner to the electron ($F_{e}$), whose present collider limit (13 TeV LHC) on its mass should be around $m_{F_{e}} \geq$ few TeV.

\end{abstract}

\section{\large Introduction.}
Very minor typos corrected to "Eur.Phys.J. C78 (2018) no.2, 150" in section 4 as follows:\\
\\
$|h_{11}|^{2} = 1.14\times 10^{-5}$ $\rightarrow$ $|h_{11}|^{2} = 10^{-7}$ and $v_{0} = 185$ GeV $\rightarrow$ $v_{0} = 1.4$ TeV and so $m_{11} = M_{11} \approx O(0.1)$ eV $\rightarrow$ $m_{11} = M_{11} \approx O(0.05)$ eV.\\
\\
The self consistency of the standard-model (SM) has been established (tentatively at least) with LHC discovery of the scalar behaving like the Higgs scalar of the SM. And about years ago since the SM has been formulated, it has went through numerous experimental test and no major disagreement with its prediction has been discovered. But still our universe turn out to be at least little more complicated than the SM can anticipate. One major discovery that point towards incompleteness of SM is the presence of dark-matter (DM). Another was the discovery of neutrino oscillation which means that neutrinos has small masses where as in SM neutrinos are massless. Then also it turn out that CP violation in SM due to Kobayashi-Maskawa (KM) \cite{KM} phase turn out to be too small to explain the observed excess of matter over the anti-matter. So the most important questions that theoretical and experimental efforts in years to come will be related to nature of DM, mechanism behind the neutrino oscillation phenomena and search for new sources of CP violations.\\
In this work we will extend the SM by adding one more $U(1)_{F}$ guage group to it with introducing only new exotic scalars and leptons. This simple extension turn out to explain the DM relic density, loop generated neutrino masses, Baryon genesis and muon (g-2) anomaly. One of the peculiar features of this particular realization of $U(1)_{F}$ extension of SM is that, due to solutions to the axial anomaly free conditions, if the electric charge of $F_{\tau}$ is taken as a free parameter, then the electric charge of $F_{e}$ will have opposite sign to that of the $F_{\mu}$, so if $F_{\mu}$ to explain the observed deviation in muon (g-2) then $F_{e}$ is require to be stable.\\
This paper is organized as follows. In the next section we introduce various particles in our model and write down Lagrangian invariant under all the symmetries imposed on the particles. Then in section III and IV we dwell into the consequencies of the model related to muon (g-2), neutrino mass, DM, Baryon-Genesis and collider signature of the exotic charged leptons. We conclude in section V.

\section{Model details.}

In this section we give details of the model. Our model introduces one additional $U(1)_{F}$ local gauge to the SM gauge group. So the full gauge group of our model is $SU(3)_{c}\times SU(2)_{L}\times U(1)_{Y}\times U(1)_{F}$. Only new exotic leptons and extra-scalars, which are singlet under the SM gauge group $SU(3)_{c}\times SU(2)_{L}$ are introduced to the IDM particle content. We introduce three charged leptons $F_{e, \mu, \tau}$ with each new leptons carrying respective lepton number 1 i.e for instance, lepton number of $F_{e}$ is $(\mathcal{N}_{e}, \mathcal{N}_{\mu}, \mathcal{N}_{\tau}) = (1, 0, 0)$ etc. We require only the right handed component of the new leptons to be charged under the $U(1)_{F}$. These new leptons are odd under a discrete $Z_{2}$ symmetry. Also three more additional neutral right handed Majorana fermions $N_{iR}$ are added which are also odd under the $Z_{2}$ but these neutral fermions does not carry $U(1)_{F}$ charges. In addition to those new fermions, we have the one inert doublet $\eta$ which carries the same charges under the SM gauge group as the SM Higgs except that this new scalar doublet is odd under the $Z_{2}$ (Inert-doublet model (IDM)) whose VEV is zero. The new neutral fermions can have very large Majorana masses $M_{i}$ given as $\bar{N}_{iR}M_{i}N_{iR}^{c}$, which, along with $\eta_{0}$, can help generate neutrino masses at one loop \cite{Ma-model}. We introduce another scalar $\phi$, singlet under all the gauge symmetries except it is odd under the discrete $Z_{2}$. We also require one real scalar, $\phi_{1}$, singlet under the full gauge group of the model and even under $Z_{2}$ which develops a non-zero VEV $v_{1}$. Also we introduce one complex scalar, $\phi_{2}$, which is charged under the $U(1)_{F}$ and develops a non-zero VEV $v_{2}$ that breaks the $U(1)_{F}$ gauge symmetry spontaneously and give mass to the new gauge boson $Z_{F}$. We require the real scalar and the complex scalar just introduced because non-trivial solution to the anomaly free condition lead to one exotic lepton uncharged and other two exotic leptons charged under $U(1)_{F}$. So we require two scalars to generate masses for the fermions carrying different $U(1)_{F}$ charges including zero. The new particles and their charges under various symmetries are tabulated in the Table \ref{tab1}.

\begin{table}[h!]
\begin{center}
\begin{tabular}[b]{|c|c|c|c|c|c|} \hline
Particles & $SU(3)_{c}$ & $SU(2)_{L}$ & $U(1)_{Y}$ & $U(1)_{F}$ & $Z_{2}$ \\
\hline\hline
$F_{iR}$ & 1 & 1 & $Y_{i}$ & $n_{i}$ & -1 \\
\hline
$F_{iL}$ & 1 & 1 & $Y_{i}$ & 0 & -1 \\
\hline
$N_{iR}$ & 1 & 1 & 0 & 0 & -1 \\
\hline
$\phi$ & 1 & 1 & 0 & 0 & -1 \\
\hline
$\phi_{1}$ & 1 & 1 & 0 & 0 & +1 \\
\hline
$\phi_{2}$ & 1 & 1 & 0 & $n_{\phi_{2}} = n_{\mu} = -n_{e}$ & +1 \\
\hline
$\eta$ & 1 & 2 & 1/2 & 0 & -1 \\
\hline
\end{tabular}
\end{center}
\caption{The charge assignments of new particles under the full gauge group $SU(3)_{c}\times SU(2)_{L}\times U(1)_{Y}\times U(1)_{F}$. Our choice of $U(1)_{F}$ charge of $\phi_{2}$ is related to our choice of particular solution of the axial anomaly free equations. In this particular choice, $\phi_{2}$ gives masses to $F_{e}$ and $F_{\mu}$ and $\phi_{1}$ gives mass to $F_{\tau}$. Where index $i = e, \mu, \tau$ and our Y of SM $U(1)_{Y}$ is same as $\frac{Y}{2}$ in the usual convention.}
\label{tab1}
\end{table}
To make the theory free of axial anomaly the $Y_{i}$s and $n_{i}$s must solve the following equations,
\be
\begin{split}
\sum^{3}_{i = 1}Y_{i}^{2}n_{i} = 0\\
\sum^{3}_{i = 1}n_{i}^{2}Y_{i} = 0\\
\sum^{3}_{i = 1}n^{3}_{i} = 0
\end{split}
\label{anomaly1}
\ee
which are the anomaly free conditions coming from $U(1)_{Y}^{2}U(1)_{F}$, $U(1)_{F}^{2}U(1)_{Y}$ and $U(1)_{F}^{3}$ respectively. But there is one more anomaly free condition due to gravity as $Gravity^{2}U(1)_{F}$ which gives
\be
\sum^{3}_{i = 1}n_{i} = 0.
\label{anomaly2}
\ee
One simplest solutions of Eqs.(\ref{anomaly1}) and Eqs.(\ref{anomaly2}) for $Y_{i}$s and $n_{i}$s is trivial solution i.e $n_{e} = n_{\mu} = n_{\tau} = 0$. The trivial solution is same as not introducing the $U(1)_{F}$ gauge symmetry at all as all the new leptons decouples from the $Z_{F}$ gauge boson. As it will be clear in the following paragraphs, the advantage of introducing the $U(1)_{F}$ is that non-trivial solutions to the axial anomaly free conditions given above require that one of the exotic leptons is forbidden to have Yukawa coupling with its corresponding SM lepton, and therefore should not show any deviation in the (g-2) of the corresponding SM lepton. Since (g-2) of electron has shown no deviation from the SM prediction, it is clear that $y_{e}\bar{e}_{R}\phi F_{e}$ should be the forbidden one.\\
A non-trivial solution of Eqs.(\ref{anomaly1}) and Eqs.(\ref{anomaly2}) for $Y_{i}$s and $n_{i}$s that is interesting is given when we set either $n_{e}$ or $n_{\mu}$ or $n_{\tau}$ equal to zero. In this work we set $n_{\tau} = 0$ which make the electric charge of $F_{\tau}$ a free parameter. Then the four equations are solved for $Y_{e}$, $Y_{\mu}$, $n_{e}$ and $n_{\mu}$ if we set $Y_{e} = -Y_{\mu}$ and $n_{e} = -n_{\mu}$, but for $F_{\mu}$ to explain the observed anomaly in $\delta a_{\mu}$ \cite{our1}, we require $Y_{\mu} = Q_{F_{\mu}} = -1$ which implies that $Y_{e} = Q_{F_{e}} = +1$ and so charge conservation and Lorenz invariance forbid $F_{e}$ to contribute to $\delta a_{e}$, which is consistent with the experimental findings as no deviation in $\delta a_{e}$ from the SM prediction has been reported. If we require Yukawa terms such as $y_{\mu}\bar{\mu}_{R}F_{\mu L}\phi$, to explain the observed anomaly in muon (g-2), then term such $y_{e}\bar{e}_{R}F_{e L}\phi$ are forbiden by charge conservation and term such as $y_{e}\bar{e}_{R}F_{e L}^{c}\phi$ is forbiden by Lorenz invariance, so $F_{e}$ will be a stable heavy charged particle. Although $Y_{\tau}$ is a free parameter, if we also set $Y_{\tau} = Q_{F_{\tau}} = -1$, then we can expect a deviation in $\delta a_{\tau}$ in future measurements coming from $y_{\tau}\bar{\tau}_{R}\phi F_{\tau}$ term.

\subsection{Scalar sector.}

The genaral Lagrangian invariant under all the symmetries of the model in the scalar sector can be written as
\be
\mathcal{L}_{scalar} = |D^{F}_{\mu}\phi_{2}|^{2} + |D_{\mu}H|^{2} + |D_{\mu}\eta|^{2} + |\partial_{\mu}\phi|^{2} + \partial_{\mu}\phi_{1}|^{2} - V(H,,\eta,\phi,\phi_{1},\phi_{2})
\ee
where $D^{F}_{\mu} = \partial_{\mu} - in_{\phi_{2}}g_{F}Z_{F\mu}$ and $D_{\mu} = \partial_{\mu} -ig\sigma\cdot W_{\mu} - iYg^{'}B_{\mu}$ with H being the SM Higgs scalar, $\eta$ being the inert-doublet and $Z_{F\mu} $ and $B_{\mu}$ being gauge fields of $U(1)_{F}$ and $U(1)_{Y}$ respectively with $\phi$, $\phi_{1}$ and $\phi_{2}$ being new SM singlet scalars. The scalar potential can be expressed as
\be
\begin{split}
V(H,\phi,\phi_{1},\phi_{2}) = m^{2}H^{\dagger}H + m_{\eta}^{2}\eta^{\dagger}\eta + m_{\phi}^{2}\phi^{\dagger}\phi + m_{1}^{2}\phi_{1}^{\dagger}\phi_{1} + m_{2}^{2}\phi_{2}^{\dagger}\phi_{2} + \lambda_{1}(H^{\dagger}H)^{2} + \lambda_{2}(\eta^{\dagger}\eta)^{2}\\
+ m_{H\eta \phi}(H^{\dagger}\eta + \eta^{\dagger}H)\phi + \lambda_{3}(H^{\dagger}H)(\eta^{\dagger}\eta) + \lambda_{4}|H^{\dagger}\eta|^{2} + \frac{\lambda_{5}}{2}\{ (H^{\dagger}\eta)^{2} + h.c \}\\
+ \lambda_{\phi}(\phi^{\dagger}\phi)^{2} + \lambda_{\phi_{1}}(\phi_{1}^{\dagger}\phi_{1})^{2} + \lambda_{\phi_{2}}(\phi_{2}^{\dagger}\phi_{2})^{2} + \lambda_{H\phi}(H^{\dagger}H)(\phi^{\dagger}\phi)\\
+ \lambda_{H1}(H^{\dagger}H)(\phi^{\dagger}_{1}\phi_{1}) + \lambda_{H2}(H^{\dagger}H)(\phi^{\dagger}_{2}\phi_{2}) + \lambda_{\eta\phi}(\eta^{\dagger}\eta)(\phi^{\dagger}\phi) + \lambda_{\eta\phi_{1}}(\eta^{\dagger}\eta)(\phi^{\dagger}_{1}\phi_{1})\\
+ \lambda_{\eta\phi_{2}}(\eta^{\dagger}\eta)(\phi^{\dagger}_{2}\phi_{2}) + \lambda_{1\phi}(\phi^{\dagger}_{1}\phi_{1})(\phi^{\dagger}\phi) + \lambda_{2\phi}(\phi^{\dagger}\phi)(\phi^{\dagger}_{2}\phi_{2}) + \lambda_{12}(\phi^{\dagger}_{1}\phi_{1})(\phi^{\dagger}_{2}\phi_{2})
\end{split}
\label{scalar-pot}
\ee
where we have $H^{T} = (G^{+}, \frac{v_{0} + h + iG^{0}}{\sqrt{2}})$, $\eta^{T} = (H^{+}, \frac{\eta_{R} + i\eta_{I}}{\sqrt{2}})$, $\phi = \frac{\phi_{R} + i\phi_{I}}{\sqrt{2}}$, $\phi_{1} = v_{1} + \sigma$ and $\phi_{2} = \frac{v_{2} + S + iG^{s}}{\sqrt{2}}$ with $G^{\pm}$, $G^{0}$ and $G^{s}$ being the Nambu-Goldstone bosons. And $v_{0} \approx 246$ is the SM Higgs VEV and similarly $v_{1}$ and $v_{2}$ are the VEV of real field $\phi_{1}$ and the complex field $\phi_{2}$ respectively. We take the couplings in the limit such that the SM Higgs H and the inert-doublet $\eta$ decouples from the the other new singlet scalars i.e $(\lambda_{H1}, \lambda_{H2}, \lambda_{H\phi}) \approx (\lambda_{\eta \phi_{1}}, \lambda_{\eta \phi_{2}}, \lambda_{\eta \phi}) \approx 0$. Also with hindsight, to explain the muon (g-2)anomaly and DM, we require the mixing between $\eta$ and $\phi$ to be very small, and since terms containing the $m_{H\eta\phi}$ mix the $\eta$ and $\phi$ after Higgs developes a nozero VEV, we set $m_{H\eta\phi} = 0$ in this work so that no mixing between $\eta$ and $\phi$ is introduced. Then the mass of the Higgs particle is
\be
M^{2}_{h} = -2m^{2} = 2\lambda_{1}v_{0}^{2}
\ee
where as the masses of the charged Higgs $H^{\pm}$ and two neutral scalars $\eta^{0}_{R}$ and $\eta^{0}_{I}$ are given by
\be
\begin{split}
M^{2}_{H^{\pm}} = m_{\eta}^{2} + \lambda_{3}v_{0}^{2}/2,\\
M^{2}_{\eta_{R}^{0}} = m_{\eta}^{2} + (\lambda_{3} + \lambda_{4} + \lambda_{5})v_{0}^{2}/2,\\
M^{2}_{\eta_{I}^{0}} = m_{\eta}^{2} + (\lambda_{3} + \lambda_{4} - \lambda_{5})v_{0}^{2}/2.
\end{split}
\ee
As we can see from the above equation, which of the two ($\eta_{R}^{0}$ or $\eta^{0}_{I}$) is the lighter depends on the sign of the $\lambda_{5}$, if $\lambda_{5} < 0$ then $\eta_{R}$ is the lighter one and so DM candidate and vice versa.

\subsection{Fermionic sector.}

As mentioned before, the symmetries of the model allows the neutral fermions $N_{iR}$ to have very large Majorana masses given as $\bar{N}_{iR}M_{i}N_{iR}$ where $M_{i}$ is a $3\times 3$ Majorana mass matrix. We do not entertain in this work how such a very heavy Majorana masses can be generated which could come from unified theories such as supergravity etc., see \cite{Yanagita}-\cite{Fukugita-Yanagita}.  The symmetries of the model also allows Yukawa terms such as $h_{ij}\bar{L}_{i}i\sigma_{2}\eta N_{iR}$, which along with the Majorana mass terms for the $N_{iR}$ can generate neutrino masses at one loop \cite{Ma-model}. We can also have Yukawa terms such as $y_{i}\bar{l}_{iR}\phi F_{iL}$ where $l_{iR}$ 's are SM right handed charged leptons. But non trivial solutions to the $U(1)_{F}$ anomaly free conditions restrict at maximum only two of the three leptons is allowed to have such Yukawa interactions. Since no deviation has been reported in the electron (g-2) other than the SM predicted value, in the following sections we will assume that $F_{\mu L}$ and $F_{\tau L}$ have electric charge given by $Q_{F_{\mu, \tau}} = Y_{F_{\mu}, F_{\tau}} = -1$ and then the axial gauge anomaly free conditions sets $Q_{F_{e}} = Y_{F_{e}} = +1$, so then only muon and tau (g-2) has contributions from respective exotic leptons in our model. The general fermionic sector Lagrangian invariant under the full symmetries of the model can be written as
\be
\begin{split}
\mathcal{L}_{fermion} = \sum^{\tau}_{i = e}\bar{F_{i}}\gamma^{\mu}(iD_{i\mu})F_{i} + \sum^{\tau}_{i = e}\bar{F_{i}}\gamma^{\mu}(i\partial_{\mu}P_{L})F_{i} + \sum^{\tau}_{i = e}\bar{N}_{Ri}M_{i}N_{Ri}^{c}\\
+ \sum^{\tau}_{i, j = e}h_{ij}\bar{L}_{i}i\sigma_{2}\eta^{*} N_{jR} + \sum^{\tau}_{i = \mu}y_{i}\bar{l}_{iR}\phi F_{iL} + \sum^{\mu}_{i = e}g_{F_{i}}\bar{F}_{iR}\phi_{i}F_{iL} + g_{F_{\tau}}\bar{F}_{\tau R}\phi_{1}F_{\tau L} + h.c
\end{split}
\label{Ferm-Lang}
\ee
where $D_{i\mu} = \partial_{\mu}P_{R} - iY_{i}g^{'}B_{\mu} - in_{i}g_{F_{i}}Z_{F\mu}P_{R}$ with $P_{R/L} = \frac{1}{2}(1 \pm \gamma_{5})$ and $\phi^{\dagger}_{e} = \phi_{\mu} = \phi_{2}$ as $n_{e} = -n_{\mu}$ from the solutions of the anomaly free conditions. The $B_{\mu}$ is the SM $U(1)_{Y}$ gauge boson and $Z_{F\mu}$ being the new $U(1)_{F}$ gauge boson. We can express the $B_{\mu}$ in terms of SM $Z_{\mu}$ and $A_{\mu}$ as
\be
B_{\mu} = -\sin\theta_{W}Z_{\mu} + \cos\theta_{W}A_{\mu},
\ee
so then we can express the interaction of the new exotic leptons with the SM weak gauge boson $Z_{\mu}$ and photon $A_{\mu}$ as
\be
\sum_{i = e}^{\tau}\bar{F}_{i}\gamma^{\mu}Y_{i}g'B_{\mu}F_{i} = e\sum_{i = e}^{\tau}Y_{i}\bar{F}_{i}(-\tan\theta_{W}Z_{\mu} + A_{\mu})\gamma^{\mu}F_{i}
\label{EM-current}
\ee
where $g' = \frac{e}{\cos\theta_{W}}$. From the above Eqs.(\ref{EM-current}), we can see that the production of these new leptons at LHC will be via the process $pp \rightarrow Z^{*}/\gamma^{*} \rightarrow F^{+}_{i}F^{-}_{i}$. Even though the production of the three exotic leptons are similar in nature, their signature mainly coming from their decays differs drastically. This is because, since $y_{e}\bar{e}_{R}\phi F_{eL}$ term is forbidden by the axial gauge anomaly free conditions, in this model, the exotic lepton $F_{e}$ will be a stable charged particle just like its SM counterpart, the electron. The other two exotic leptons, $F_{\mu}$ and $F_{\tau}$, are not stable as they can decay via $F_{\mu/\tau} \rightarrow \mu/\tau + \phi$. Due to unbroken $Z_{2}$ symmetry, $\phi$ will be a stable particle if $m_{F_{\mu}} > m_{\phi} + m_{\mu}$ is satisfied and so $\phi$ can be a DM candidate. But we will see in section \ref{g-2}, if $y_{\mu}\bar{\mu}_{R}\phi F_{\mu L}$ to explain the muon (g-2) within 1 $\sigma$ of the experimental value which require large $y_{\mu}$, then contribution of $\phi$ to the present relic density of DM will be negligibly tiny. In early universe, most of the $F_{e}$ would have been depleted via $F_{e}^{+}F_{e}^{-} \rightarrow Z_{F}/Z^{*}/\gamma^{*} \rightarrow F_{\mu}^{+}F_{\mu}^{-}(F_{\tau}^{+}F_{\tau}^{-}) \rightarrow \mu^{+}\mu^{-}(\tau^{+}\tau^{-}) + missing\ energy\ (\phi \phi)$. The collider signature of $F_{e}$ will be basically the collider signature of a very heavy stable charged particle, and given LHC reaching 13 TeV reported no such signatures of a stable heavy charged particle, we expect the mass of the $F_{e}$ to be larger than few TeV. However for the other two exotic leptons, the main collider signatures will be
\be
e^{+}e^{-}/pp \rightarrow Z^{*}/\gamma^{*} \rightarrow F_{\mu}^{+}F^{-}_{\mu}(F_{\tau}^{+}F^{-}_{\tau}) \rightarrow \mu^{+}\mu^{-}(\tau^{+}\tau^{-}) + missing\ energy\ (\phi \phi),
\ee
and other key collider signatures will be
\be
e^{+}e^{-}/pp \rightarrow Z^{*}/\gamma^{*} \rightarrow \gamma + \phi_{1}/\phi_{2} (via\ triangle\ loop) \rightarrow \gamma + missing\ energy\ (\phi_{1}/\phi_{2} \rightarrow \phi \phi)
\ee
and
\be
e^{+}e^{-}/pp \rightarrow Z^{*}/\gamma^{*} \rightarrow \gamma + \phi_{1}/\phi_{2} (via\ triangle\ loop) \rightarrow \gamma + (\gamma + \gamma(or\ Z))(via\ triangle\ loop).
\ee
In this sense we think $e^{+}e^{-}$ colliders such as ILC (being a precision machine) will be more sensitive in detecting the missing energies in the final states of above reactions than Hadron colliders such as LHC.

\section{Muon (g-2) anomaly.}
\label{g-2}

The contribution from the term $y_{\mu}\bar{\mu}_{R}\phi F_{\mu L}$ from Eqs.(\ref{Ferm-Lang}) to the muon (g-2) anomaly can be expressed as \cite{Leveille}
\be
\delta a_{\mu} = \frac{m^{2}_{\mu}y^{2}_{\mu}}{16\pi^{2}}\int^{1}_{0}dx\frac{x^{2} - x^{3}}{m^{2}_{\mu}x^{2} + (m^{2}_{F_{\mu}} - m^{2}_{\mu})x + m^{2}_{\phi}(1 - x)}.
\label{muong2}
\ee
And as pointed out in \cite{our1}, in this type of models where the scalar in the loop does not develops VEV, the constrains from the heavy charged lepton searches, which has rule out at $90\%$ C.L for $m_{F_{Heavy}} \le 100.8 GeV$, does not apply because the heavy charged leptons in our model can not decay into the final state those experiments looked for such as $F^{\pm} \rightarrow W^{\pm}\nu$ and $F^{\pm} \rightarrow l^{\pm}Z$. Therefore in our model the new charged lepton ($F_{\mu}$) can have mass less the 100 GeV, but the mass of $F_{\mu}$ must satisfy $m_{F_{\mu}} > m_{Z}/2$ so that no contribution from $F_{\mu}$ to the Z decay width (via $Z \rightarrow \bar{F_{\mu}}F_{\mu}$), which is very precisely measured and no deviation from SM prediction has been found. Similarly if $m_{\phi} < m_{Z}/2$ then also contribution to the invisible Z decay width ($Z \rightarrow \phi\phi$ via triangle loop) is expected to be large and so we require $m_{\phi} > m_{Z}/2$. In agreement with the experimental non-observation of long lived heavy charged particles, we require $m_{F_{\mu}} \geq m_{\phi} + m_{\mu}$ with large Yukawa coupling, i.e $y_{\mu} \approx 1$ so that $F_{\mu}$ decays quickly after production as $F_{\mu} \rightarrow \phi + \mu$.\\
All these constrains on the mass can be satisfied along with explaining the muon (g-2) anomaly within $1\sigma$ of the $\delta a_{\mu}^{Exp}$ if we have
\be
\frac{y_{\mu}}{m_{F_{\mu}}} \approx 0.0188 GeV^{-1},
\ee
where we have taken the approximation of $m_{\phi} \approx m_{F_{\mu}} >> m_{\mu}$ in the Eqs.(\ref{muong2}). Then for $y_{\mu} \approx 1$, we have $m_{\phi} \approx m_{F_{\mu}} \approx 53$ GeV, and for $y_{\mu} \approx 2\sqrt{\pi}$, we have $m_{\phi} \approx m_{F_{\mu}} \approx 189$ GeV, assuming the condition $m_{F_{\mu}} > m_{\phi} + m_{\mu}$ is satisfied. Due to unbroken $Z_{2}$ symmetry, $\phi$ will be a DM candidate, but as pointed out in \cite{Abe}, for such large Yukawa couplings the contributions to the present relic density of DM by $\phi$ will be negligibly tiny. 
In case of fermion $F_{\mu}$ affecting the Z decay $Z \rightarrow \bar{\mu}\mu$ via the triangle loop, by adapting the formula II.39 of \cite{Chaing}, we have from the triangle loop contribution to $Z \rightarrow \bar{\mu}\mu$ given as
\be
Br(Z \rightarrow \bar{\mu}\mu)_{triangle} = \frac{G_{F}}{3\sqrt{2}\pi}\frac{m_{Z}^{3}}{(16\pi^{2})^{2}\Gamma_{Z}}s_{w}^{2}t_{w}^{2}y_{\mu}^{4}|F_{2}(F_{\mu},\phi) + F_{3}(F_{\mu},\phi)|^{2}
\ee
where $t_{w} = \frac{s_{w}}{\sqrt{1-s^{2}_{w}}}$ and $s_{w} = \sin\theta_{w}$ with $\theta_{w}$ being the Wienberg angle and
\be
\begin{split}
F_{2}(F_{\mu},\phi) = \int^{1}_{0}dx(1-x)\ln[(1-x)m^{2}_{F_{\mu}} + xm^{2}_{\phi}],\\
F_{3}(F_{\mu},\phi) = \int^{1}_{0}dx\int^{1-x}_{0}dy\frac{(xy-1)m^{2}_{Z} + (m^{2}_{F_{\mu}}-m^{2}_{\phi})(1-x-y) - \Delta\ln \Delta}{\Delta}
\end{split}
\ee
with $\Delta = -xym^{2}_{Z} + (x+y)(m^{2}_{F_{\mu}}-m^{2}_{\phi}) + m^{2}_{\phi}$ and total Z width $\Gamma_{Z} = 2.4952 \pm 0.0023$ GeV. For $m_{F_{\mu}} \approx 53 \approx m_{\phi}$ and $y_{\mu} = 1$, we have $Br(Z \rightarrow \bar{\mu}\mu)_{triangle} = 2.19 \times 10^{-6}$ compared to the $Br(Z \rightarrow \bar{\mu}\mu)_{Exp} = (3.366 \pm 0.007)\%$ \cite{Patrignani}, so the triangle loop contribution from the new particles to $Br(Z \rightarrow \bar{\mu}\mu)$ is an order of magnitude smaller than the error in the experimental value.

\section{Loop generation of neutrino masses and Dark-Matter.}

\begin{figure}[h!]
\begin{minipage}[t]{0.48\textwidth}
\hspace{0.4cm}
\includegraphics[width=2\linewidth, height=8cm]{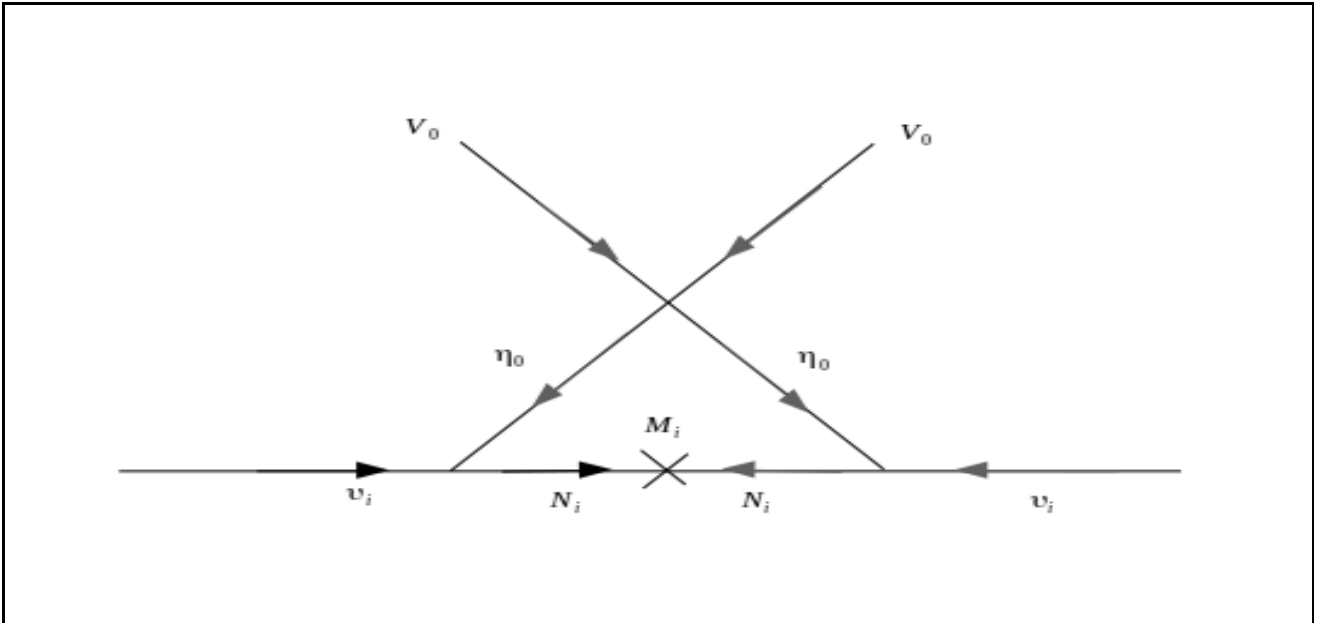}
\end{minipage}
\caption{Neutrino mass generation at one loop level. Where here $V_{0}$ is the VEV of SM Higgs and $\eta_{0}$ denoting either $\eta_{R}$ or $\eta_{I}$.}
\label{Fig1:fig2}
\end{figure}

It has been shown more than ten years ago in \cite{Ma-model2}, that with addition of three neutral Majoran fermions which are also odd under the $Z_{2}$ symmetry to the inert-doublet model (IDM), the origins of small neutrino mass, DM and Baryogenesis can all be linked and explained. The small Majorana neutrino masses for the left handed neutrinos can arise at one loop level via the scotogenic mechanism \cite{Ma-model} as shown in Figure \ref{Fig1:fig2}, where the neutral components of the inert-doublet and $N_{i}$ propagating in the loop inducing indirect interaction of the left handed neutrinos with the Higgs vacuum and $M_{i}$ and generate a mass term $M_{\alpha \beta}\bar{\nu}^{c}_{\alpha}\nu_{\beta}$. In the scotogenic model the loop generated neutrino mass matrix $M_{\alpha \beta}$ can be expressed as \cite{Ma-model}
\be
M_{\alpha \beta} = \sum_{i}\frac{h_{\alpha i}h_{\beta i}M_{i}}{16\pi^{2}}[ \frac{m_{R}^{2}}{m^{2}_{R} - M^{2}_{i}}\ln\frac{m^{2}_{R}}{M^{2}_{i}} - \frac{m_{I}^{2}}{m^{2}_{I} - M^{2}_{i}}\ln\frac{m^{2}_{I}}{M^{2}_{i}} ],
\label{Neut-mass}
\ee
where denoting the inert-doublet as $\eta^{T} = (\eta^{+}, \eta^{0})$ with $\eta^{0} = (\eta^{0}_{R} + i\eta^{0}_{I})/\sqrt{2}$ 
and $m_{R, I}$ being the masses of the the $\eta^{0}_{R, I}$ respectively. The lightest one among the $N_{i}$ and $\eta^{0}_{R, I}$ is the DM. In this work we have taken the $\eta^{0}_{R}$ as the lightest and therefore DM candidate. However as shown in \cite{Ma-model2}, if one of the $\eta^{0}_{R, I}$ is the DM then there is also the possibility of generating Baryon asymmetry by converting the lepton-asymmetry into baryon asymmetry via sphalerons \cite{Ma-ref12}. As a numerical estimate, for $|h_{11}|^{2} \approx 1\times 10^{-7}$ and the term in the bracket in Eqs.(\ref{Neut-mass}) being equal to $\frac{v_{0}^{2}}{M_{1}^{2}}$, where $N_{R1}$ being the lightest of the heavy neutrinos and $v_{0} \approx 1.4$ TeV, then for $M_{1} \approx 2.6\times 10^{7}$ GeV we get the masses of the light neutrinos as $m_{11} = M_{11} \approx O(0.05)$ eV and also these parameters will be able to generate required baryon excess \cite{Fukugita-Yanagita}. Also in this model constrains from small neutrino masses has little effect on determining the DM annihilation crossection and therefore determining the DM relic density.

The neutral Lightest Stable Particle (LSP) of the IDM can be a DM candidate has been first pointed out in \cite{Ma-model}. We take $\lambda_{5} < 0$ and so of the two neutral scalars $\eta_{R}$ and $\eta_{I}$, $\eta_{R}$ is the DM candidate in our model. In \cite{Okada-74}\cite{Okada-75}, before the Higgs discovery at LHC, it has been shown that in detail analysis $\eta_{R}$ can account the present relic density of DM if the masses of the scalar DM lies in two regions, $m_{\eta_{R}} \le m_{W} \approx 80$ Gev or $m_{\eta_{R}} \geq 500$ Gev assuming the Higgs mass as 120 GeV or 200 GeV. However, in \cite{Goudelis-2015}\cite{Goudelis-2013} the authors did a thorough analysis of IDM DM after discovery of the SM like Higgs at $m_{h} \approx 126$ GeV, combine with direct searches and LHC Run One data, they have ruled out the low and intermediate mass region of IDM DM except in a very narrow strip near the $m_{\eta_{R}} = m_{h}/2$, corresponding to very small values of the $h-\eta_{R}-\eta_{R}$ coupling (the funnel regions). The recent LUX \cite{LUX} and PandaX \cite{PandaX} data reaffirm their conclusions more strongly. Therefore we will focus on the high mass region where the DM mass is $m_{\eta_{R}} \geq 500$ GeV. In this region if $m_{\eta} < m_{\eta_{R}}$, then the annihilation into two gauge bosons dominate, and if $m_{\eta_{R}} > m_{\eta}$, then annihilation into two Higgs bosons dominate. It turn out coannihilation plays little role in this region. In the limit of small mass splitting and $\lambda_{L} \approx 0$, the relic density can be accounted for $m_{\eta_{R}} \geq 500$ GeV \cite{Okada-75} where $\lambda_{L} = (\lambda_{3} + \lambda_{4} + \lambda_{5})/2$. The contribution to relic density tends to increase as $m_{\eta_{R}}$ increase but this can be controlled by increasing the mass splitting as annihilation crossection tends to increase as mass splitting increases \cite{Okada-74}\cite{Okada-75}. However, in DM theories (including ours) where it is assumed that in some point in the early universe DM particles are in thermal equilibrium with the rest of the matter, their annihilation crossection is bound by the partial waves unitarity of the S-matrix, which in turn constrains the relic density and mass of the particle. The unitarity bounds on total annihilation crossection for a scalar DM particle is given as \cite{Honorez-32}
\be
\left\langle \sigma v \right\rangle_{v \rightarrow 0}^{unitary} \approx \frac{4\pi}{m^{2}_{\eta_{R}}}\sqrt{\frac{x_{f}}{\pi}},
\ee
where $x_{F} = \frac{m_{\eta_{R}}}{T_{f}}$ with $T_{f}$ being the freeze out temperature. It has been shown in \cite{Honorez-32} that both unitarity and WMAP constrains can be satisfied for the scalar DM mass $m_{\eta_{R}} \le$ 130 TeV.

\section{Conclusions.}

In this work we have proposed a simple extension of the SM by only adding three $Z_{2}$ odd exotic charged leptons ($F_{e},\ F_{\mu},\ F_{\tau}$) whose right handed component are charged under a new $U(1)_{F}$ gauge symmetry, three $Z_{2}$ odd neutral fermions ($N_{1},\ N_{2},\ N_{3}$) singlet under the entire gauge symmetry, one $SU(2)_{L}$ doublet ($\eta$) odd under a discrete $Z_{2}$ symmetry whose VEV is zero (the inert-doublet model), one $Z_{2}$ odd scalar ($\phi$) singlet under the entire gauge group and whose VEV is zero, one $Z_{2}$ even scalar ($\phi_{1}$) also singlet under the entire gauge group which develops a non zero VEV $v_{1}$, one more $Z_{2}$ even scalar ($\phi_{2}$) charged under the $U(1)_{F}$ whose non-zero VEV breaks the $U(1)_{F}$ gauge symmetry spontaneously and give mass to the gauge boson $Z_{F}$. With addition of the above new particles, we have been able to build a model which can give two DM candidate in terms of lightest neutral component of the inert-doublet ($\eta_{R}$ in our case) and $\phi$ but if $\phi$ to explain the muon (g-2) anomaly than $\phi$ will contribute only a tiny fraction to the present DM relic density, so most of the present relic density of DM will consist of $\eta_{R}$ whose mass can be in the range 500 GeV to 130 TeV. In this model neutrino masses are generated at one loop level as well as baryon-genesis via lepto-genesis is also possible. We have also given key signatures of these new exotic leptons at LHC and future $e^{+}e^{-}$ colliders. We find the the key signature will be in the form of $e^{+}e^{-}/pp \rightarrow Z^{*}\gamma^{*} \rightarrow F_{\mu}^{+}F^{-}_{\mu}(F_{\tau}^{+}F^{-}_{\tau}) \rightarrow \mu^{+}\mu^{-}(\tau^{+}\tau^{-}) + missing\ energy\ (\phi \phi)$ and $e^{+}e^{-}/pp \rightarrow Z^{*}\gamma^{*} \rightarrow \gamma + \phi_{2} (via\ triangle\ loop) \rightarrow \gamma + missing\ energy\ (\phi_{2} \rightarrow \phi \phi)$. One of the most peculiar signature of this model is the existence of a stable very heavy (about $m_{F_{e}} \geq$ few TeV) charged lepton partner to the electron.


{\large Acknowledgments: \large} This work is supported and funded by the Department of Atomic Energy of the Government of India and by the Government of U.P.

\end{document}